\documentclass[draft]{agujournal2019}
\usepackage{url} %this package should fix any errors with URLs in refs.
\usepackage[inline]{trackchanges} %for better track changes. finalnew option will compile document with changes incorporated.
\usepackage{soul}
\draftfalse

\journalname{Geophysical Research Letters}

\begin{document}

%% ------------------------------------------------------------------------ %%
%  Title
%
% (A title should be specific, informative, and brief. Use
% abbreviations only if they are defined in the abstract. Titles that
% start with general keywords then specific terms are optimized in
% searches)
%
%% ------------------------------------------------------------------------ %%

% Example: \title{This is a test title}

%\title{LBT SHARK-VIS Observes a Major Resurfacing Event on Io}
\title{Observation of Io’s Resurfacing via Plume Deposition Using Ground-based Adaptive Optics at Visible Wavelengths with LBT SHARK-VIS}

%% ------------------------------------------------------------------------ %%
%
%  AUTHORS AND AFFILIATIONS
%
%% ------------------------------------------------------------------------ %%

% Authors are individuals who have significantly contributed to the
% research and preparation of the article. Group authors are allowed, if
% each author in the group is separately identified in an appendix.)

% List authors by first name or initial followed by last name and
% separated by commas. Use \affil{} to number affiliations, and
% \thanks{} for author notes.
% Additional author notes should be indicated with \thanks{} (for
% example, for current addresses).

% Example: \authors{A. B. Author\affil{1}\thanks{Current address, Antartica}, B. C. Author\affil{2,3}, and D. E.
% Author\affil{3,4}\thanks{Also funded by Monsanto.}}

% \authors{=list all authors here=}

\authors{ Al Conrad\affil{1},
          Fernando Pedichini\affil{2,3},
          Gianluca Li Causi\affil{2,3},
          Simone Antoniucci\affil{2,3},
          Imke de Pater\affil{4},
          Ashley Gerard Davies\affil{5},
          Katherine de Kleer\affil{6},
          Roberto Piazzesi\affil{2,3},
          Vincenzo Testa\affil{2,3},
          Piero Vaccari\affil{2,3},
          Martina Vicinanza\affil{2},
          Jennifer Power\affil{1},
          Steve Ertel\affil{7,1},
          Joseph C. Shields\affil{1},
          Sam Ragland\affil{1},
          Fabrizio Giorgi\affil{2,3},
          Stuart M. Jefferies\affil{8},
          Douglas Hope\affil{9},
          Jason Perry\affil{10},
          David A. Williams\affil{11},
          David M. Nelson\affil{11}
}

\affiliation{1}{Large Binocular Telescope Observatory, The University of Arizona, 933 North Cherry Ave, Tucson, AZ 85721, USA}
\affiliation{2}{INAF Osservatorio Astronomico di Roma,
Via Frascati 33, 00078, Monte Porzio Catone, Italy}
\affiliation{3}{INAF-ADONI, Adaptive Optics National Laboratory, Italy}
\affiliation{4}{University of California - Berkeley,
501 Campbell Hall, Berkeley, California 94720 USA}
\affiliation{5}{Jet Propulsion Laboratory-California
   Institute of Technology, 4800 Oak Grove Drive, 
   Pasadena, CA 91109 USA}
\affiliation{6}{California Institute of Technology,
   1200 E. California Blvd.,
   Pasadena, CA 91125 USA}
\affiliation{7}{Department of Astronomy and Steward Observatory, The University of Arizona, 933 North Cherry Ave, Tucson, AZ 85721, USA}
\affiliation{8}{Department of Physics and Astronomy, Georgia State University, 25 Park Place, Atlanta GA 30303, USA}
\affiliation{9}{Georgia Tech Research Institute, 925 Dalney St., Atlanta GA 30332 USA}
\affiliation{10}{University of Arizona,
   1200 E University Blvd, Tucson, AZ 85721 USA}
\affiliation{11}{Arizona State University,
   1151 S Forest Ave, Tempe, AZ USA}

% \affiliation{1}{First Affiliation}
% \affiliation{2}{Second Affiliation}
% \affiliation{3}{Third Affiliation}
% \affiliation{4}{Fourth Affiliation}

% \affiliation{=number=}{=Affiliation Address=}
%(repeat as many times as is necessary)

%% Corresponding Author:
% Corresponding author mailing address and e-mail address:

% (include name and email addresses of the corresponding author.  More
% than one corresponding author is allowed in this LaTeX file and for
% publication; but only one corresponding author is allowed in our
% editorial system.)

% Example: \correspondingauthor{First and Last Name}{email@address.edu}

\correspondingauthor{Albert Conrad}{aconrad@lbto.org}

%% Keypoints, final entry on title page.

%  List up to three key points (at least one is required)
%  Key Points summarize the main points and conclusions of the article
%  Each must be 140 characters or fewer with no special characters or punctuation and must be complete sentences

% Example:
% \begin{keypoints}
% \item	List up to three key points (at least one is required)
% \item	Key Points summarize the main points and conclusions of the article
% \item	Each must be 140 characters or fewer with no special characters or punctuation and must be complete sentences
% \end{keypoints}

\begin{keypoints}
\item High resolution images taken with SHARK-VIS at LBT reveal low and high albedo features obscuring a portion of Pele's red ring on Io.
\item This new eruption deposit likely originated from a powerful eruption in August 2021 located at Pillan Patera.
\item Such images provide a new imaging capability that yields vital context to other observations of planetary surfaces.
\end{keypoints}

%% ------------------------------------------------------------------------ %%
%
%  ABSTRACT and PLAIN LANGUAGE SUMMARY
%
% A good Abstract will begin with a short description of the problem
% being addressed, briefly describe the new data or analyses, then
% briefly states the main conclusion(s) and how they are supported and
% uncertainties.

% The Plain Language Summary should be written for a broad audience,
% including journalists and the science-interested public, that will not have 
% a background in your field.
%
% A Plain Language Summary is required in GRL, JGR: Planets, JGR: Biogeosciences,
% JGR: Oceans, G-Cubed, Reviews of Geophysics, and JAMES.
% see http://sharingscience.agu.org/creating-plain-language-summary/)
%
%% ------------------------------------------------------------------------ %%

%% \begin{abstract} starts the second page

\begin{abstract}
Since volcanic activity was first discovered on Io from \textit{Voyager} images in 1979, changes on Io's surface have been monitored from both spacecraft and ground-based telescopes.  Here, we present the highest spatial resolution images of Io ever obtained from a ground-based telescope.  These images, acquired by the SHARK-VIS instrument on the Large Binocular Telescope, show evidence of a major resurfacing event on Io’s trailing hemisphere.  When compared to the most recent spacecraft images, the SHARK-VIS images show that a plume deposit from a powerful eruption at Pillan Patera has covered part of the long-lived Pele plume deposit.  Although this type of resurfacing event may be common on Io, few have been detected due to the rarity of spacecraft visits and the previously low spatial resolution available from Earth-based telescopes.  The SHARK-VIS instrument ushers in a new era of high resolution imaging of Io's surface using adaptive optics at visible wavelengths.

\end{abstract}

\section*{Plain Language Summary}
A new instrument, called SHARK-VIS, on the Large Binocular Telescope in Arizona, has obtained high spatial resolution, visible wavelength images of Io, the highly volcanic moon of Jupiter.  Large multicolored plume deposits were imaged, revealing where the red deposit from a volcano named Pele was covered by another plume deposit from another volcano, named Pillan Patera, the site of a powerful eruption in 2021.  SHARK-VIS ushers in a new age in planetary imaging.

%% ------------------------------------------------------------------------ %%
%
%  TEXT
%
%% ------------------------------------------------------------------------ %%

\section{Introduction}

Io is a highly volcanic world which sometimes exhibits large-scale
surface changes \cite{davies07,depater2021,lopes23}. Io's volcanism is driven by strong tidal heating, induced by the Laplace (4:2:1) orbital
resonance among Io, Europa, and Ganymede \cite{peale1979}. That Io is currently volcanically active was discovered by the imaging of giant volcanic plumes of gas and dust by \textit{Voyager 1} \cite{morabito1979, smith1979}. Subsequently, hundreds of active volcanoes have been identified via thermal emission observed in the infrared, and through the presence of surface changes, 
like plume deposits, at visible wavelengths.  Plume deposits around the larger 
volcanoes can produce large-scale, multicolored, albedo changes that dominate the local landscape.  On occasion, it is only the presence of the deposits that reveal that an eruption had taken place.  The large Pele-type plumes \cite{mcewen83} have a high exit velocity ($>$1 km/s), can reach heights of over 300 km, and are difficult to image directly. The deposits from this plume type are annular, can be over 1000 km in diameter, and are red in color, the result of the presence of short-chain (S$_3$ and S$_4$) sulfur allotropes. As these allotropes change to yellow cyclo-octal (S$_8$) sulfur under Io surface conditions in a few months, these large plume deposits are therefore ephemeral, as has been seen at Surt (335\textdegree W, 42\textdegree N) and South of Karei (13\textdegree W, 12\textdegree S) \cite{geissler2004, davies07}.

An exception to the short-lived nature of the giant plume deposits is Pele.  First observed by \textit{Voyager 1} in 1979 \cite{smith1979}, Pele's plume deposit is continually being renewed by degassing from a persistent lava lake \cite{davies2001} at 255.6\textdegree W, 18.4\textdegree S, a location on Io's trailing hemisphere. The resulting plume deposit is over 1200 km across. Close to the center of the plume deposit, around the lava lake, darker material is deposited, likely clasts of silicate lava caught in the gas stream which detrain from the gas flow as the gases expand. The Pele lava lake at the center of the plume deposit is a long-lived thermal source, although it has diminished in strength over the past decade \cite{davies2001,depater2016,dekleer2023}. Whether the amount of material being ejected has also diminished in strength has been an unanswered question as the last time this area of Io was imaged at visible wavelengths was in 2007 by \textit{New Horizons} (\cite{spencer2007} (see also NASA Photojournal image PIA09355).

Notably, in 1997 during the \textit{Galileo} mission, a dark, likely silicate-rich deposit with a relatively narrow, higher-albedo aureole originating from a nearby eruption at Pillan (245\textdegree W, 11\textdegree S) covered a portion of Pele's red ring.
% as shown in Figure \ref{galFig}.
Over the next four years, instruments on \textit{Galileo} watched as the thermal emission from the Pillan eruption waned \cite{davies2001} and the dark Pillan deposit was mostly erased by the continuing deposition of material from the Pele plume \cite{davies2001,keszthelyi01,turtle04}. 

Now, over 25 years later, a similar resurfacing event has been detected
by SHARK-VIS \cite{pedichini22} at the Large Binocular Telescope (LBT),
% On UT 2023 November 23 and 2024 January 10, SHARK-VIS,
one of the first visible light, diffraction limited imagers
operating on an 8-10 meter telescope in the Northern Hemisphere.
% acquired high resolution images of Io's trailing
% hemisphere that reveal this resurfacing event.

\section{Observations}

\subsection{Adaptive Optics at Visible Wavelengths}

Adaptive Optics (AO), the technology that employs a deformable mirror
to compensate for atmospheric distortion, has been in use on large
telescopes for over 25 years \cite{lai97}.  Until recently, however, the science
path for the system has been restricted almost exclusively to infrared wavelengths.
Thanks largely to two emerging technologies, high density deformable
mirrors and fast visible light detectors with low read out noise, 
a higher resolution era is emerging for large telescopes
equipped with imagers optimized for visible AO.

While there will always be a need for infrared images of Io, particularly
at L (3.8 $\mu$m) and at M (4.8 $\mu$m) bands, to detect thermal emission
from active volcanoes, visible-light images are essential to ``see'' the landscape, i.e., identify locations of eruptions and associated features such as plume deposits. As shown in this paper, such images 
can now be obtained at high spatial resolution from ground-based telescopes
on Earth with a spatial resolution of approximately 80km at 550 nm. Hitherto visible-light images could only be obtained by spacecraft, or with the Hubble Space Telescope (HST);
%\cite{spencer1997}; the spatial 
the spatial
resolution of the latter images are, however, roughly a factor of 3 lower than the images obtained with SHARK-VIS on the LBT, due to the difference in aperture size.
%When using an 8 meter class telescope (e.g., one side of the Large Binocular
%Telescope) we achieve a resolution of 25-30 mas from our ground based
%observation, approximately 50km on the surface of Io.  Diffraction limit at 500 %nm??

\subsection{SHARK-VIS on the LBT}

SHARK-VIS saw first light at LBT on October 2nd, 2023.
The instrument shares the center-bent Gregorian focus on the
right-hand (DX) side of the LBT with the Large Binocular Telescope
Interferometer \cite{hinz2016, ertel2020}, where
it picks up the visible light from the direct beam using a deployable dichroic.
The remaining red-visible and infrared light is transmitted to
the LBTI and the shared wavefront sensor \cite{bailey2014}
which has recently received the Single-conjugated adaptive Optics
Upgrade for the LBT
\cite{pinna_SOUL}.
% The SOUL system takes advantage of a higher frame rate and an 
% improved EMCCD detector to provide higher Strehl ratio than its predecessor, 
% FLAO \cite{esposito_FLAO}.
SHARK-VIS is an imager operating in the 400-1000 nm 
wavelength range and delivers spatial resolutions down to $\sim$15 mas 
(at 550 nm), which in the infrared bands will be achieved only by future extremely large telescopes (ELTs).
With a plate scale of 6.5 mas/pixel, SHARK-VIS provides 
slight oversampling of the Nyquist frequency for this resolution.

The spatial resolution given above ($\sim$15 mas at 550 nm),
is computed as follows:
The entrance pupil of the AO system reduces the 8.4 meter LBT aperture
to 8.25 meters.  The full width half max (FWHM) of the Airy disk is then
$1.03 \times \lambda / D = 14.2 (\sim$15) mas for $\lambda$ = 550 nm.

The LBT AO system delivers this level of performance in ideal atmospheric conditions
for targets within $40^{\circ}$ of zenith (1.3 airmass or less).
For the observations reported here, the achieved spatial resolution was
$\sim$24 mas (as described in section \ref{sec:mfbd} below).

% allowing for advanced image 
% post processing techniques as MFBD and myopic deconvolution thanks to the SOUL 
% telemetry data as demonstrated by using the SHARK-VIS pathfinder ForeRunner
% \cite{Hope22}.
SHARK-VIS is equipped with an Andor Zyla 4.2 PLUS sCMOS science camera that can operate at a 
frame rate up to 1 Khz with a relatively low readout noise; this makes it possible to employ a 
fast imaging approach that helps freeze both 
atmospheric turbulence and telescope vibrations and allows for both post-facto registration and advanced processing of the frames with a significant improvement of the image sharpness \cite{pedichini17, stangalini17, Hope22}. SHARK-VIS has a wide-band atmospheric dispersion corrector (ADC) enabling the use of 150nm wide-band 
filters from 500 to 1000 nm and 50 nm wide bands in the range 400-500 nm while 
covering zenith angles up to 40 degrees. 

\subsection{Acquisition and data reduction}
\label{sec:acq}

\begin{figure}
 \noindent\includegraphics[width=\textwidth]{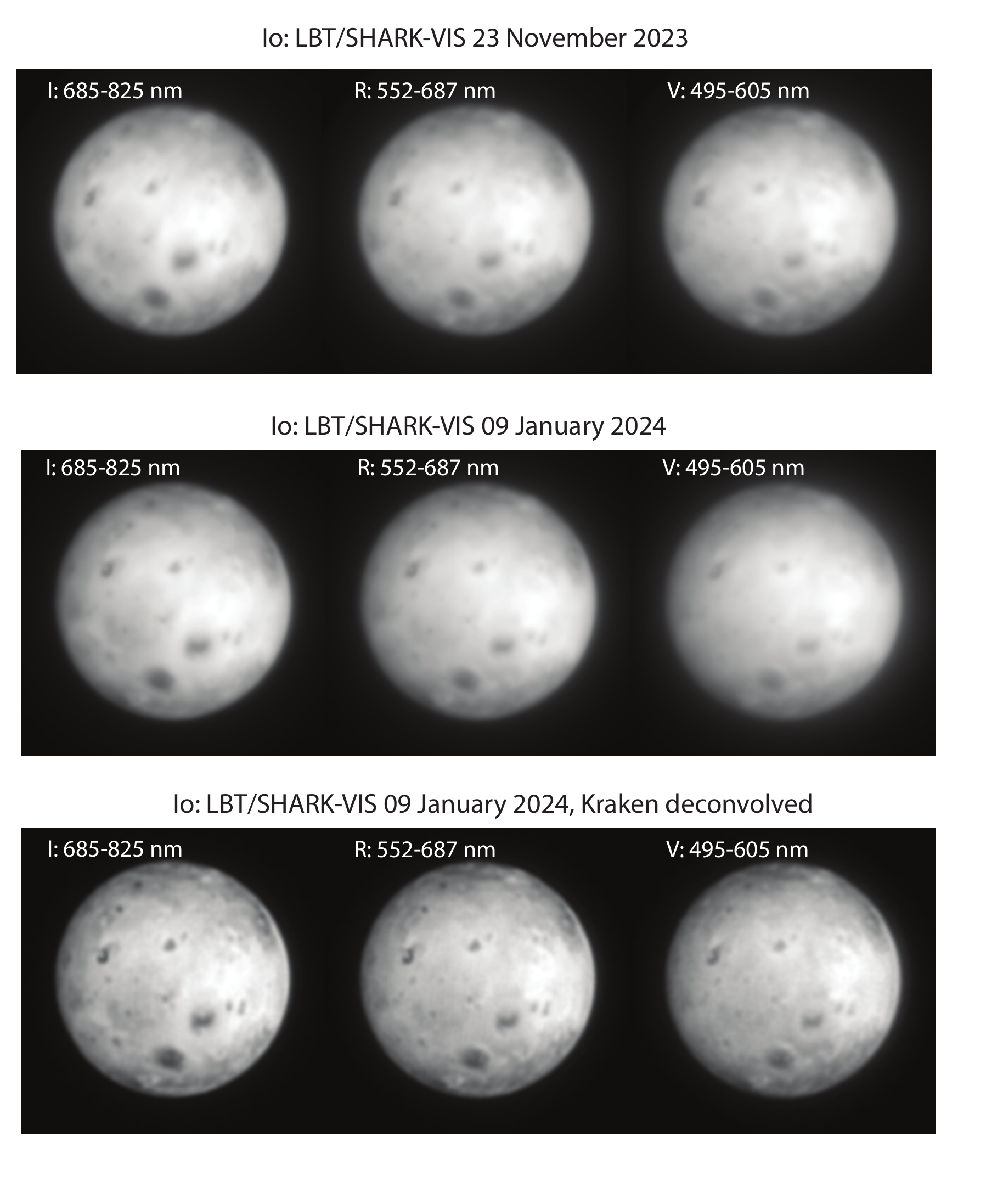}
 \caption{Images of Io obtained with SHARK-VIS on the LBT in November 2023 and January 2024. The images are re-scaled to Io's apparent diameter. The third row shows the Kraken deconvolved images from January 2024. The filter used is shown on each image. Io North is approximately up.}
 \label{LBT_Shark_IRV}
\end{figure}

We observed Io with SHARK-VIS on UT November 23rd, 2023
and UT January 10th, 2024, taking data in each of 3 filters, as tabulated
(together with the relevant ephemerides)
in Table \ref{tab:obs}.
Both November and January observations consisted of a sequence of two minutes of short exposures for each filter (10 ms and 5 ms, respectively, which were corrected for darks and the flat field. During processing, the frames were ranked for sharpness, based on the maximum spatial frequency above the noise level in the power spectra; the 1000 best frames were selected, whose average UT is reported in Table \ref{tab:obs}.
The frames were then de-rotated to sky orientation by
correcting for the parallactic angle (due to the fact that SHARK-VIS does not have a mechanical de-rotator). As a final step, the center of the body was measured in each frame by fitting a circle to the object’s limb, found as the maximum gradient locus in all directions from the image baricenter outwards, and the frames were co-registered via Fourier shifting, then average stacked, and finally oriented to have Io's North pole pointing upwards, based on JPL’s HORIZONS ephemerides (\url{https://ssd.jpl.nasa.gov/horizons/app.html#/}).
The resulting images are shown in the first two rows of Fig.\ref{LBT_Shark_IRV}. The January images are more detailed, even though the apparent diameter of Io was larger in November. This is because, due to a technical problem,
data collected during the November run suffered from an imperfect correction of non-common-path aberration by the AO system.
The three filters were also combined to form the tri-color images shown in Fig. \ref{Io_Shark_Vis}, where the I,R,V bands have been rendered as RGB respectively. Image stretching and saturation have been slightly enhanced to improve the visibility of faint morphological structures on the satellite's surface.
As a result of this processing, the colors produced are different from the colors seen in \textit{Galileo} images (Fig. \ref{Io_Shark_Vis} top center panel) and Voyager images, which were also not true color.
The annular Pele plume deposit, however, still appears red in the SHARK-VIS images since it is less absorbing in the I-band (Fig. \ref{LBT_Shark_IRV}).

\subsection{Multi-frame blind deconvolution}
\label{sec:mfbd}
A 5 millisecond frame sequence like that acquired from SHARK-VIS contains
more information than can be retrieved by simple shift-and-add stacking.
In fact, the frame-to-frame PSF diversity can be exploited to jointly estimate the instantaneous wavefront and reconstruct the brightness distribution of the object at a nearly diffraction limited resolution.
To realize this approach,
we applied Kraken multi-frame blind deconvolution \cite{Leist2024, Hope22} to the 250 sharpest frames of the January acquisition, which produced a much more detailed view of Io in all the three bands (Fig. \ref{LBT_Shark_IRV} bottom panel),
and in the corresponding color composite
(Fig. \ref{Io_Shark_Vis} bottom left frame), at a FWHM resolution of 3.63 pixels (i.e. 24 mas for the plate scale of SHARK-VIS), as computed from the maximum frequency of its power spectrum.
% which corresponds to a spatial resolution of 80 km on the body surface at the moment of the observation
For Io's distance in January (see Table \ref{tab:obs}), a single SHARK-VIS
pixel spanned $\sim$22 km.
The achieved spatial resolution on the surface of Io
is therefore $\sim$80 km.

\begin{table}
    \centering
    \caption{Observing parameters for the LBT SHARK-VIS observation of Io}
    \begin{tabular}{cccccc}
         &&&&\\
         UT & filter & $\lambda$ (nm) & \# of frames$^c$ &
           \multicolumn{2}{c}{exposure time} \\
         &&&& total (s) & per frame (ms) \\
         \hline
         23 November 2023$^a$: \\ 
         7h 11m 16.5s & V & 495 - 605 & 1000 & 10.0 & 10 \\
         7h 13m 40.5s & I & 685 - 825 & 1000 & 10.0 & 10 \\
         7h 16m 00.5s & R & 552 - 687 & 1000 & 10.0 & 10 \\
         10 January 2024$^b$: \\  
         2h 31m 23:4s & V & 495 - 605 &  1000 &   5.0 &  5 \\
         2h 30m 48:3s & I & 685 - 825 &  1000 &   5.0 &  5 \\
         2h 30m 18:1s & R & 552 - 687 &  1000 &   5.0 &  5 \\
         \hline
    \end{tabular}

%    \caption{Observing parameters for the LBT SHARK-VIS observation of Io}
%    \begin{tabnote}
    
$^a$ The observer-target ($\Delta$) and heliocentric ($r$) distances; Io's diameter ($D$); 
the sub-observer W. longitude ($\lambda_E$) and latitude ($\theta_E$); and the sub-solar W. longitude ($\lambda_S$) and sub-solar latitude ($\theta_S$) are: 

$\Delta =$ 4.052 AU, $r =$ 4.977 AU, $D =$ 1.24'', $\lambda_E\sim$264$^\circ$, $\theta_E =$ 3.30$^\circ$,  $\lambda_S \sim$260$^\circ$, $\theta_S =$3.13$^\circ$. 

$^b$ $\Delta =$ 4.619 AU, $r =$ 4.986 AU, $D =$ 1.09'', $\lambda_E\sim$273$^\circ$, $\theta_E =$ 3.03$^\circ$,  $\lambda_S \sim$262$^\circ$, $\theta_S =$3.14$^\circ$. 

$^c$ Number of selected frames (using the criteria given in section \ref{sec:acq}).  The total number of frames collected were 12000 and 24000 for 23 November and 10 January, respectively.\\

The Ephemerides are obtained from JPL's HORIZONS system: \url{https://ssd.jpl.nasa.gov/horizons/app.html#/}
%\end{tabnote}

    \label{tab:obs}
\end{table}

\section{Results: A resurfacing event at Pillan Patera}

\begin{figure}
 \noindent\includegraphics[width=\textwidth]{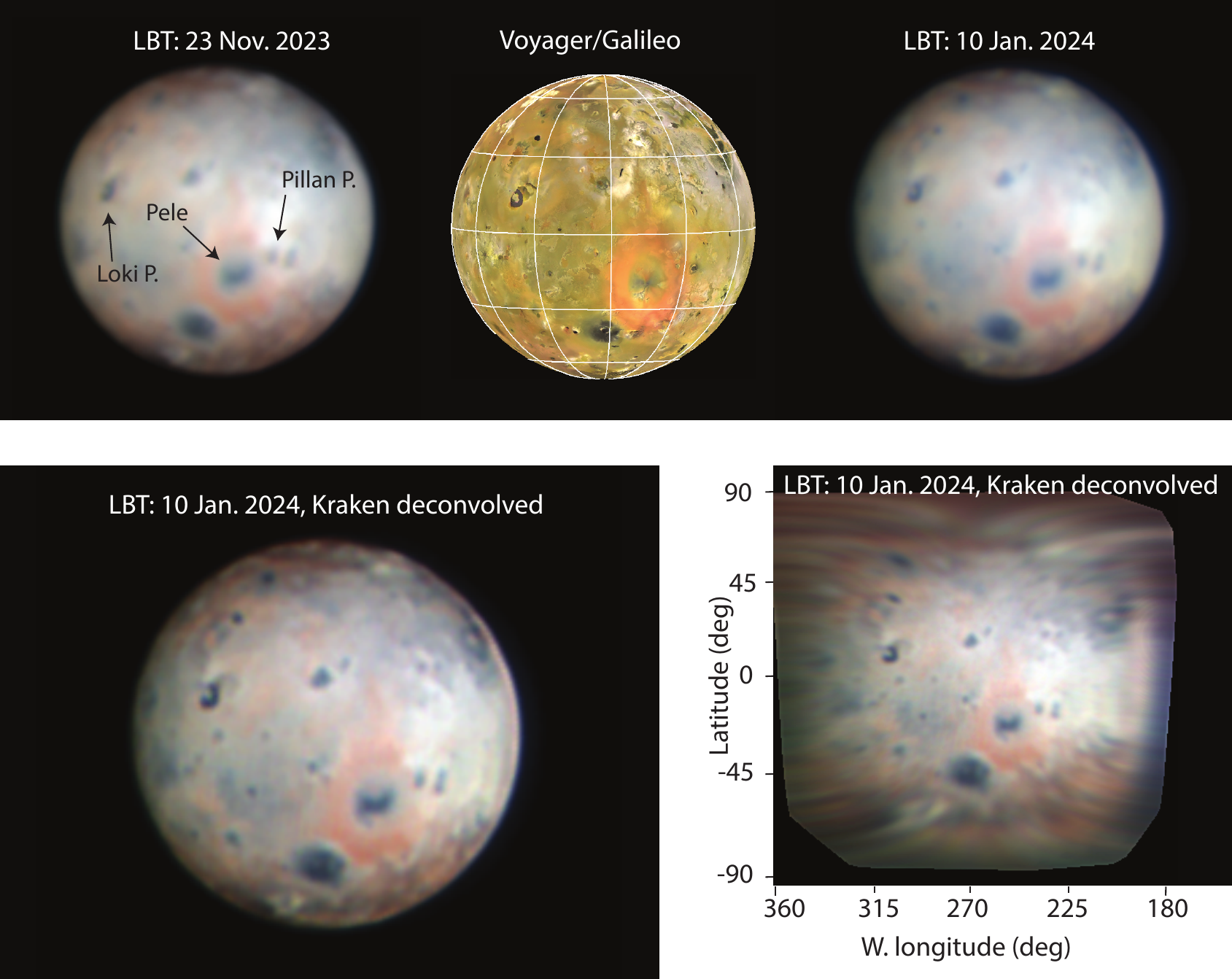}
 \caption{The SHARK-VIS detection image on Nov. 23, 2023 (upper left), on Jan. 10, 2024 (upper right),
   and the reprojection of the \textit{Voyager} and \textit{Galileo} spacecraft-derived Io photomosaic for Jan. 10, 2024 (center) \cite{becker05}. Labels to Pele, Pillan Patera and Loki Patera are shown on the upper left image. The bottom row shows the Kraken-deconvolved image from Jan. 10, 2024 on the left; the bottom right map shows this image reprojected onto equirectangular (approximate) longitude-latitude coordinates. All images have been rotated so Io North is up.
   No detectable change to the Pillan plume deposit has occurred during the 50 days separating the two
   observations.}
 \label{Io_Shark_Vis}
\end{figure}

As seen in the left- and right-hand panels of Fig. \ref{Io_Shark_Vis}, and with more details in the bottom panels,
the red annular deposit centered on Pele is missing a segment.  About ¼ of the ring is obscured by a high albedo deposit with a low albedo center.  The bright deposit is likely SO$_2$ ice, which is ubiquitous on Io and a common component of volcanic plumes and plume deposits. 
The previous most recent high resolution visible imaging of the area
was performed during an Io fly-by by the \textit{New Horizons} spacecraft in 2007 (see above), at which time the ring was intact. Additionally, there is a dark area centered on nearby Reiden Patera, another volcano active during the \textit{Galileo} and \textit{Juno} epochs \cite{Veeder2015IoHF, davies24}.  Another dark area is seen at Ra Patera, the site of an extensive resurfacing event imaged by \textit{Galileo} \cite{geissler2004}. Ra Patera was imaged by JunoCam during \textit{Juno} orbit PJ58 on 2 March 2024 (e.g., image \\
JNCE\_2024034\_58C00026\_V01). The patera exhibits a reddish rim and dark floor with associated extra-patera dark flows.  Taken together, these dark features are consistent with the SHARK-VIS imagery.  Such low albedo areas are expected from recent or ongoing silicate volcanism. 

\begin{figure}
 \noindent\includegraphics[width=\textwidth]{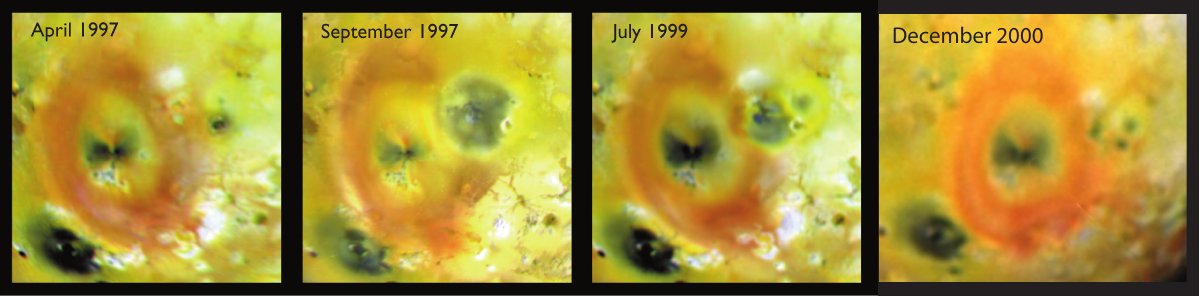}
 \caption{
Four images following the 1997 Pillan eruption which show the plume
deposit covering a portion of Pele's red ring for over 2 years.
April 1997 (left): pre-eruption. September 1997 (middle left). The eruption may have started in May but only peaked in June \cite{davies2001}, and then steadily decreased in thermal emission over the next few years. July 1999 (middle right):
the Pele plume deposits had begun burying the Pillan deposits, but there is still evidence of an interaction between the Pele plume and degassing from the Pillan eruption site and/or emplaced flows. December 2000 (\textit{Galileo} orbit 29): the Pele plume has completely recovered. (Image credits: both NASA/JPL/University of Arizona.  The three left panels are from  \url{https://photojournal.jpl.nasa.gov/catalog/PIA02501}. The right panel is from \url{https://photojournal.jpl.nasa.gov/catalog/PIA02588}).
}
\label{galFig}
\end{figure}

The SHARK-VIS images of Pele and Pillan Patera are reminiscent of an image obtained in September 1997 by the Solid State Imaging experiment (SSI) \cite{belton92} on the \textit{Galileo} spacecraft, which showed that Pele's ring was partly covered by a plume deposit from a powerful, voluminous eruption originating to the northwest of Pillan Patera (see %table \ref{tline97} and 
Fig. \ref{galFig})
 \cite{geissler2004}.
The Pele deposit had partially recovered by July 1999, slightly less than two years after the Pillan deposit formed. By late December 2000 the Pele plume deposit had completely recovered \cite{turtle04}.

 %Other large red plume deposits similar to Pele are emplaced quickly %and can fade rapidly \cite{davies07} when there is no renewable %source of sulphur. A giant ring deposit from an unnamed patera south %of Karei (13\textdegree W, 12\textdegree S) was emplaced around 31 %May 1998 but had faded by August 14 1999  \cite{geissler2004}.

%The most likely source of the material that is covering the Pele plume %deposit as seen in the SHARK-VIS images is another powerful eruption, %this time emanating from Pillan Patera (244\textdegree W, %12\textdegree S, as indicated in figure \ref{Io_Shark_Vis} upper %right panel)
%that took place during August 2021,
%rather than from the extra-patera surface as in 1997 %\cite{keszthelyi01}. 

The most likely source of the material that is covering the Pele plume deposit as seen in the SHARK-VIS images is another powerful eruption near Pillan Patera (244\textdegree W, 12\textdegree S, as indicated in Fig. \ref{Io_Shark_Vis} upper left panel). In the following text we examine when such an eruption was most likely to have taken place.

\begin{figure}
  \noindent\includegraphics[width=\textwidth]{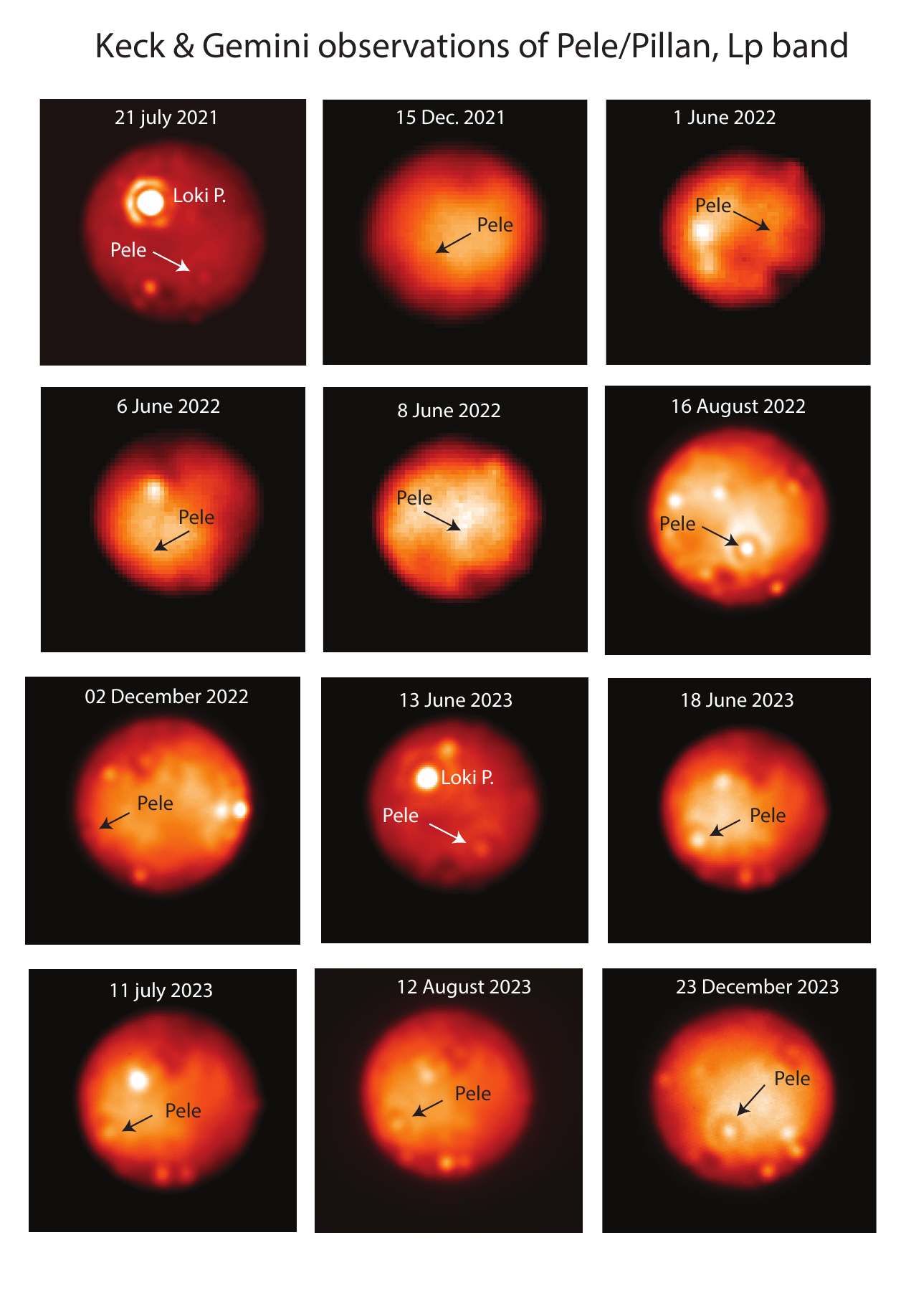}
  \caption{Keck and Gemini images of Io in Lp band (3.8 $\mu$m) of Pele/Pillan, which show no emission at Pillan Patera or the immediate surrounding area. Images from 15 December 2021 and from June 2022 are taken with the Gemini telescope. All other images are taken with the Keck telescope. Images from 2 Dec. 2022, 13 and 18 June 2023 are from the Keck Twilight Zone website https://www2.keck.hawaii.edu/inst/tda/TwilightZone.html. All images have Io North up, and Pele has been indicated.}
  \label{fig:Keck_Io}
\end{figure}

Examination of observations of Io in the infrared from different observation systems, both telescopes 
(Gemini, Keck, IRTF) and spacecraft (\textit{Juno}), allow a timeline of events to be compiled. Table \ref{tab:timeline} provides an eruption history for Pillan Patera spanning three years prior to, and just beyond, the SHARK-VIS detection. 
Keck and Gemini images documenting the non-detections in Table \ref{tab:timeline} are shown in Fig. \ref{fig:Keck_Io}; the upper limit values are based upon prior observations with these instruments
%\cite{dekleer2016,depater2016}.
\cite{depater2016}.
Recent \textit{Juno} JIRAM infrared data analysis shows 4.8-$\mu$m thermal emission from an active, volcanic source, albeit at a relatively low level (average 4.46 GW/$\mu$m), emanating from Pillan Patera \cite{davies24}. There are no other thermal sources detected by JIRAM between 2017 and early 2023 close to the location of the black deposit identified in Fig. \ref{Io_Shark_Vis}. On August 13th, 2021, an extremely powerful eruption, designated as an "outburst" event, was detected at or near Pillan Patera by the InfraRed Telescope Facility (IRTF) \cite{tate23}. This eruption is the most likely source of the material that covers the Pele ring, and that is seen in the SHARK-VIS images two years later.  

 The IRTF images were obtained in L and M bands, where the thermal hot spot emission clearly stands out above the solar reflected light. Two weeks later, on August 27, the emission had subsided considerably \cite{tate23}, while three weeks earlier, on July 21, no emission was detected from Pillan Patera by the Keck telescope (see Fig. \ref{fig:Keck_Io}).  The event was not observed at its height by other groundbased telescopes nor by instruments on \textit{Juno}, but is bracketed by both JIRAM and Keck observations.
Also, the S+ torus and Na nebula seem quiescent during this event \cite{Morganthaler24}.
The event in the Pillan Patera vicinity was powerful, but short-lived. 

JunoCam, the visible wavelength imager on the \textit{Juno} spacecraft \cite{hansen2017}, has obtained a number of images of Io, mostly from a high-latitude, sub-spacecraft vantage point, and at moderate- to high-phase angles. \cite{ravine2024}. Observation opportunities are necessarily limited by the spacecraft trajectories, with \textit{Juno} in a highly-inclined orbit around Jupiter. While JunoCam has returned high spatial resolution images of mainly Io’s northern
hemisphere (https://www.missionjuno.swri.edu/junocam), it had not, at the time of the SHARK-VIS observations, imaged all of the Pele
plume deposit. Coverage was limited to the western and northern portions of the plume deposit. As discussed below, on 9 April 2024 (PJ60) JunoCam did capture the entire Pele deposit; the
combination of coverage from Juno and SHARK-VIS proving to be highly complementary.

\section{Discussion}

The resurfacing scenario that occurred 25 years ago was likely similar to what took place recently, with some important changes. As seen with SHARK-VIS about two years after the Pillan outburst, the Pele plume deposit was not repaired as it was after the 1997 Pillan eruption.  We note that the plume deposit restoration timeline depends on the effusive volcanic activity within Pillan Patera and the recent rate of degassing, the level of volcanic activity at Pele, and the continuation of sulfur supply to the Pele plume.

Assuming that the Pillan Patera plume deposit was likely initially laid down in August 2021 by the outburst eruption (see Table \ref{tab:timeline}), the Pillan Patera eruption deposits persisted for at least 29 months (to January 2024). The plume deposit laid down in 1997 had mostly been buried after two years \cite{turtle04}. That the new deposit was apparently longer lasting than the 1997 deposit might be the result of less volcanic activity at Pele at present than during the \textit{Galileo} era.  The drop in activity manifested as a $\sim$75 percent drop in total thermal emission after the \textit{Galileo} mission from ~280 GW \cite{davies2001} to ~60 GW \cite{depater2016,dekleer2023,davies23a,rathbun14}.  A lower volumetric lava effusion rate might also suggest a similar decrease in gas emission volume \cite{davies07}. The plume resurfacing rate at Pele during the \textit{Galileo} epoch was estimated at 0.6 mm/yr \cite{zhang03, davies07}, and may for recent years have been only a quarter of this. 

However, a four-fold increase in thermal output from Pele was detected by JIRAM on 16 May 2023 during the \textit{Juno} PJ51 flyby of Io.  Both L-band and M-band spectral radiance from Pele exceeded 39 GW/${\mu}m$, similar to spectral radiance observed during the \textit{Galileo} epoch, and suggesting a total thermal emission of 294 GW. Activity as seen by Keck and Gemini (Fig. \ref{fig:Keck_Io}) then dropped, with Keck measuring a L-band spectral radiance of 5.2$\pm$0.6 GW/$\mu$m on 13 June 2023, and 7$\pm$0.8 GW/$\mu$m on 18 June 2023, with a 
M-band spectral radiance of 10.7$\pm {1.5}$ GW/${\mu}m$ on the latter date. These numbers are typical for all Keck observations in Figure \ref{fig:Keck_Io}. Even with this apparent resurgence of activity at Pele (albeit temporary), the plume deposit was not restored in its entirety by the time of the SHARK-VIS observations.
 
If the deposits imaged by SHARK-VIS around Pillan Patera were indeed emplaced in 2021, then their persistence reveals a complex interplay of volcanic processes. As noted above, around the dark Pillan Patera deposit in the SHARK-VIS images is a bright, high-albedo deposit (Fig. \ref{Io_Shark_Vis}), likely SO$_2$ frost. If the Io surface here is cold enough for SO$_2$ to condense it is also cold enough for plume sulfur to condense. As the rest of Pele plume deposit is seen, the plume must also be depositing material here, but it is not repairing the red ring. This suggests the presence of ongoing SO$_2$ deposition (the SO$_2$ originating from the Pillan Patera eruption) which is burying the Pele plume deposit faster than it is being emplaced, thus maintaining the gap in the ring seen by SHARK-VIS. The low albedo deposit centered on Pillan Patera persists probably because the deposition from Pele is landing on a warm, silicate-rich surface and is not condensing.  

By early April 2024, however, the Pele ring appears to have restored itself entirely.  Observations obtained by JunoCam on 9 April 2024 during orbit PJ60,
albeit at high emission angle (e.g., https://www.missionjuno.swri.edu/Vault/VaultOutput?VaultID=\\
51967\&ts=1709753150), 
cover just enough of Pele's deposit to suggest that the annular ring was once again complete. If this is the case, either SO$_2$ deposition from Pillan Patera significantly decreased or ceased entirely, or sulfur deposition from Pele increased enough to overwhelm any Pillan Patera deposition, allowing repair of the contiguous Pele plume deposit. The combination of coverage from \textit{Juno} and SHARK-VIS are highly complementary. Without the SHARK-VIS images, this resurfacing event would never have been detected. With SHARK-VIS we now have the capability to regularly monitor the evolving plume deposits at Pele and Pillan Patera for years to come, as well as other such events on Io's surface. 
SHARK-VIS is available to researchers affiliated with one or more of the LBT member institutions (listed at the end of the \textit{Acknowledgements} section below) and to other potential users via director's discretionary time

\section{Conclusions}

SHARK-VIS ushers in a new era in planetary imaging. On its very first use imaging Io, SHARK-VIS observed a major change of volcanic origin on Io's surface, providing vital visual wavelength context for the interpretation of infrared and other observations by extending contemporaneous coverage of Io. With SHARK-VIS, we now have the capability to monitor how Io's surface changes and apply what is observed to better understand volcanic activity at Pele, Pillan, and elsewhere, with implications for the manner and rate of the resurfacing of Io.

%%% Suggested section heads:
% \section{Introduction}
%
% The main text should start with an introduction. Except for short
% manuscripts (such as comments and replies), the text should be divided
% into sections, each with its own heading.

% Headings should be sentence fragments and do not begin with a
% lowercase letter or number. Examples of good headings are:

% \section{Materials and Methods}
% Here is text on Materials and Methods.
%
% \subsection{A descriptive heading about methods}
% More about Methods.
%
% \section{Data} (Or section title might be a descriptive heading about data)
%
% \section{Results} (Or section title might be a descriptive heading about the
% results)
%
% \section{Conclusions}

%Text here ===>>>

%%

%  Numbered lines in equations:
%  To add line numbers to lines in equations,
%  \begin{linenomath*}
%  \begin{equation}
%  \end{equation}
%  \end{linenomath*}

%% Enter Figures and Tables near as possible to where they are first mentioned:
%
% DO NOT USE \psfrag or \subfigure commands.
%
% Figure captions go below the figure.
% Table titles go above tables;  other caption information
%  should be placed in last line of the table, using
% \multicolumn2l{$^a$ This is a table note.}
%
%----------------
% EXAMPLE FIGURES
%
% \begin{figure}
% \includegraphics{example.png}
% \caption{caption}
% \end{figure}
%
% Giving latex a width will help it to scale the figure properly. A simple trick is to use \textwidth. Try this if large figures run off the side of the page.

%\begin{figure}
% \noindent\includegraphics[width=\textwidth]{Figures/rowOf3.png}
% \caption{The SHARK-VIS detection image (left), again with labels (right),
%   and the reprojection of \textit{Voyager} and \textit{Galileo} spacecraft imaging (center).}
% \label{rowOf3}
%\end{figure}

%\begin{figure}
% \noindent\includegraphics[width=\textwidth]{Figures/galFig.png}
% \caption{
%Three images following the 1997 Pillan eruption which show the plume
%deposit covering a portion of Pele's red ring.
%April 1997 (left): pre-eruption (which we think started in May but only peaked in June – Davies et al., 2001 - and then steadily decreased in thermal emission over the next few years). September 1997 (middle). July 1999 (right):
%the Pele plume deposits had begun burying the Pillan deposits.  In this
%right hand image there is still evidence of interaction between the Pele
%plume and degassing from the Pillan eruption site and/or emplaced flows.
%These are likely still boiling off surface volatiles.
%}
% \label{galFig}
%\end{figure}

\begin{table}
    \centering
    \begin{tabular}{cccc}
     Pillan, spectral radiance & Pele, spectral radiance & Date & \textit{Juno} orbit, or telescope \\
     M-band, GW/${\mu}m$ & M-band, GW/${\mu}m$ & Year month day\\
         \hline
        2.191             & 10.9        & 2021 Feb 21  &  PJ32i    \\
        3.364             & 11.7        & 2021 Apr 15  &  PJ33     \\
        $<$3              & 8.7$\pm$1.5    & 2021 Jul 21  &  Keck     \\
        527 $\pm$ 118 L-band &         & 2021 Aug 13  &  IRTF     \\
        Faint L-band, some M-band & & 2021 Aug 27  &  IRTF     \\
        $>$23.371        & 9.9            & 2021 Oct 16  &  PJ37i    \\
        $<$9 L-band      & $<$9 L-band    & 2021 Dec 15  &  Gemini-N \\
        16.537           & 10         & 2022 Feb 24  &  PJ40     \\
        $<$9 L-band     & $<$9 L-band     & 2022 Jun 01  &  Gemini-N \\
        $<$9 L-band     & $<$9 L-band     & 2022 Jun 06  &  Gemini-N \\
        $<$9 L-band     & $<$9 L-band     & 2022 Jun 08  &  Gemini-N \\
        8 $\pm$ 1       & 10$\pm$1.5      & 2022 Aug 16  &  Keck     \\
        $<$3            & Faint, on limb  & 2022 Dec 02  &  Keck     \\
        2.097           &         & 2023 Mar 01  &  PJ49     \\
        0.882           & 44          & 2023 May 16  &  PJ51     \\
        $<$4 L-band     & 5.2$\pm$0.6 L-band   & 2023 Jun 13  &  Keck     \\        
        $<$3            &10.4$\pm$1.5      & 2023 Jun 18  &  Keck     \\
        $<$3            &14$\pm$2      & 2023 Jul 11  &  Keck     \\
        $<$3            &11$\pm$1.4      & 2023 Aug 12  &  Keck     \\
        Visible wavelengths &      & 2023 Nov 23  &  LBT/SHARK-VIS \\
        $<$3            &10$\pm$2      & 2023 Dec 23  &  Keck     \\
        Visible wavelengths &      & 2024 Jan 24  &  LBT/SHARK-VIS \\
        JunoCam           &        & 2024 Apr 09  &  PJ60 \\
        \hline
        &&\\
    \end{tabular}
    \caption{
Observations of the Pillan Patera and Pele regions from early 2021 through early 2024, encompassing the major 2021 eruption at Pillan Patera and the 2023-24 SHARK-VIS observations.
IRTF-derived spectral radiances are from \cite{tate23}.
Note that the positions in \cite{tate23} are $\pm{10}^{\circ}$.
The thermal emission estimate for 2021 Aug 13 has been corrected
for emission angle and converted to GW/$\mu$m. All Keck observations have been corrected for the emission angle.
The data from \textit{Juno} JIRAM are from \cite{davies24}.}
\label{tab:timeline}
\end{table}

\newpage

\section*{Open Research}

The data for this observation are recorded in FITS files as follows.
These data are available from Zenodo \cite{svIo2024}.
\begin{itemize}
\item For the November observation: 2023\_11\_22-IRV\_centered.fits
\item For the January observation: 2024\_01\_09-IRV\_centered.fits
\end{itemize}

\acknowledgments
Observations have benefited from the use of
ALTA Center (alta.arcetri.inaf.it) forecasts
performed with the Astro-Meso-Nh model.
Initialization data of the ALTA automatic
forecast system come from the General
Circulation Model (HRES) of the European
Centre for Medium Range Weather Forecasts. The authors are grateful to Marco Stangalini for his help during the development of the Forerunner of SHARK-VIS and his suggestions about the fast imaging approach with sCMOS detectors. The Kraken code was developed under the Air Force Office of Scientific Research award number FA9550-14-1-0178. We thank the Keck Observatory staff for their dedication to obtaining Twilght Zone images. We acknowledge the vital work of the \textit{Juno} Project in obtaining Io data. Part of this work was performed at the Jet Propulsion Laboratory – California Institute of Technology, under Government contract.  Ashley Davies, Jason Perry, David Williams and David Nelson are supported by the NASA New Frontiers Data Analysis Program (NFDAP) under award 80NM0018F0612. The LBT is an international collaboration among institutions in the United States, Italy and Germany. LBT Corporation Members are: The University of Arizona on behalf of the Arizona Board of Regents; Istituto Nazionale di Astrofisica, Italy; LBT Beteiligungsgesellschaft, Germany, representing the Max-Planck Society, The Leibniz Institute for Astrophysics Potsdam, and Heidelberg University; The Ohio State University, representing OSU, University of Notre Dame, University of Minnesota and University of Virginia.

%% ------------------------------------------------------------------------ %%
%% References and Citations

%%%%%%%%%%%%%%%%%%%%%%%%%%%%%%%%%%%%%%%%%%%%%%%
%
\bibliography{svIo}
%
% don't specify bibliographystyle

% In the References section, cite the data/software described in the Availability Statement (this includes primary and processed data used for your research). For details on data/software citation as well as examples, see the Data & Software Citation section of the Data & Software for Authors guidance
% https://www.agu.org/Publish-with-AGU/Publish/Author-Resources/Data-and-Software-for-Authors#citation

%%%%%%%%%%%%%%%%%%%%%%%%%%%%%%%%%%%%%%%%%%%%%%%

%\bibliography{enter your bibtex bibliography filename here}

%Reference citation instructions and examples:
%
% Please use ONLY \cite and \citeA for reference citations.
% \cite for parenthetical references
% ...as shown in recent studies (Simpson et al., 2019)
% \citeA for in-text citations
% ...Simpson et al. (2019) have shown...
%
%
%...as shown by \citeA{jskilby}.
%...as shown by \citeA{lewin76}, \citeA{carson86}, \citeA{bartoldy02}, and \citeA{rinaldi03}.
%...has been shown \cite{jskilbye}.
%...has been shown \cite{lewin76,carson86,bartoldy02,rinaldi03}.
%... \cite <i.e.>[]{lewin76,carson86,bartoldy02,rinaldi03}.
%...has been shown by \cite <e.g.,>[and others]{lewin76}.
%
% apacite uses < > for prenotes and [ ] for postnotes
% DO NOT use other cite commands (e.g., \citet, \citep, \citeyear, \citealp, etc.).
% \nocite is okay to use to add references from your Supporting Information
%

\end{document}